\documentclass[aps,prb,floatfix,superscriptaddress,preprint]{revtex4-1}
\usepackage[utf8]{inputenc}
\usepackage[T1]{fontenc}
\usepackage{amssymb}
\usepackage{bm}
\usepackage{graphicx}
\usepackage{textgreek}
\usepackage[separate-uncertainty=true]{siunitx}
\usepackage[version=4]{mhchem}
\usepackage[hidelinks]{hyperref}
\usepackage[capitalize,nameinlink]{cleveref}

\usepackage{booktabs}

\newcommand{\NPS}{\texorpdfstring{\ce{NiPS3{}}}{NiPS3}}
\newcommand{\NPSPy}{\texorpdfstring{\ce{NiPS3{}/Py}}{NiPS3/Py}}
\newcommand{\Py}{\texorpdfstring{\ce{Py}}{Py}}
\newcommand{\PtPy}{\texorpdfstring{\ce{Pt/Py}}{Pt/Py}}

\newcommand{\opfl}{out-of-plane field-like}
\newcommand{\ipdl}{in-plane damping-like}

\newcommand{\tauS}{\ensuremath{\tau_\mathrm{DL}}}
\newcommand{\tauA}{\ensuremath{\tau_\mathrm{FL}}}
\newcommand{\sigmaS}{\ensuremath{\sigma_\mathrm{DL}}}
\newcommand{\sigmaA}{\ensuremath{\sigma_\mathrm{FL}}}

\DeclareSIUnit\sq{\ensuremath{\Box}}

\begin{document}
\title{
    Large interfacial spin-orbit torques in
    \texorpdfstring{\\}{ }
    layered antiferromagnetic insulator \NPS{}/ferromagnet bilayers
}
\date{\today}

\author{C. F. Schippers}
\email{c.f.schippers@tue.nl}
\affiliation{Department of Applied Physics, Eindhoven University of Technology, P.O. Box 513, 5600 MB, Eindhoven, the Netherlands}

\author{H. J. M. Swagten}
\affiliation{Department of Applied Physics, Eindhoven University of Technology, P.O. Box 513, 5600 MB, Eindhoven, the Netherlands}

\author{M. H. D. Guimarães}
\email{m.h.guimaraes@rug.nl}
\affiliation{Department of Applied Physics, Eindhoven University of Technology, P.O. Box 513, 5600 MB, Eindhoven, the Netherlands}
\affiliation{Zernike Institute for Advanced Materials, University of Groningen, P.O. Box 221, 9747 AG, Groningen, the Netherlands}

\begin{abstract}
    Finding efficient ways of manipulating magnetic bits is one of the core goals in spintronic research.
    Electrically-generated spin-orbit torques (SOTs) are good candidates for this and the search for materials capable of generating highly-efficient SOTs has gained a lot of traction in the recent years.
    While antiferromagnet/ferromagnet bilayer structures have been employed extensively for passive applications, e.g.\ by using exchange bias fields, their active properties are not yet widely employed.
    Here we show the presence of large interfacial SOTs in bilayer of a ferromagnet and the two-dimensional layered antiferromagnetic insulator \NPS{}.
    We observe a large \ipdl{} interfacial torque, showing a torque conductivity of $\sigmaS{} \approx \SI{1e5}{(\frac{\hbar}{2e}) \per(\ohm\meter)}$ even at room temperature, comparable to the best devices reported in the literature for standard heavy-metal-based and topological insulators-based devices.
    Additionally, our devices also show an \opfl{} torque arising from the \NPS{}/ferromagnet interface, further indicating the presence of an interfacial spin-orbit coupling in our structures.
    Temperature-dependent measurements reveal an increase of the SOTs with a decreasing temperature below the N\'eel temperature of \NPS{} ($T_N \approx \SI{170}{\kelvin}$), pointing to a possible effect of the magnetic ordering on our measured SOTs.
    Our findings show the potential of antiferromagnetic insulators and two-dimensional materials for future spintronic applications.
\end{abstract}

\maketitle

\section{Introduction}
The electrical manipulation of magnetization is a promising approach for novel non-volatile and energy efficient memory devices.
An especially efficient approach uses current-induced spin-orbit torques (SOTs) \cite{Liu2011,Manchon2015}, where an electric current flows through a material with high spin-orbit coupling, which applies a torque on an interfaced magnetic material.
These torques can arise from bulk effects, such as the spin-Hall effect, \cite{Haazen2013,VandenBrink2016} where the electrons in a charge current flowing through a conducting layer get deflected to opposite directions depending on their spin.
This is the main mechanism for current-induced SOTs in heavy metal/ferromagnet bilayer structures, such as Pt/Permalloy (\ce{Ni80Fe20}; \Py{}) \cite{Liu2011}.
Interfacial effects, such as the Rashba-Edelstein Effect \cite{Edelstein1990,Chernyshov2009}, can also generate a sizeable charge-to-spin conversion and can be used for SOT generation \cite{Amin2016,Amin2016a,Hellman2017}.
More recently, it has been shown that when SOTs are generated in metallic ferromagnetic \cite{Das2018,Gibbons2018,Safranski2019} and antiferromagnetic materials \cite{Zhou2019,Zhang2016a,Tshitoyan2015}, the magnetic ordering can be used to control the direction and magnitude of the generated SOTs \cite{Davidson2020}.
Even though magnetic insulators have been investigated extensively for the generation of spin currents via spin-pumping \cite{Saitoh2006,Heinrich2011} and spin Seebeck effects \cite{Uchida2010,Bauer2012} the use of antiferromagnet insulators in spin-orbit torque devices remains vastly unexplored.

\NPS{} is a layered semiconducting antiferromagnetic van der Waals crystal with a N\'eel transition temperature of approximately \SI{170}{\kelvin} in its bulk form \cite{Chittari2016}.
Below the transition temperature the magnetic moments of the hexagonally-arranged Ni atoms align in a zigzag fashion, where the coupling is ferromagnetic along a zigzag line and antiferromagnetic across it \cite{Wildes2015}.
Due to its semiconducting nature and relatively flat band dispersion, \NPS{} presents a very high resistivity unless heavily doped or under ultraviolet (UV) light illumination \cite{Foot1980,Chu2017}.
Moreover, \NPS{} also presents promising efficient catalytic properties for hydrogen evolution reaction.
Therefore, its main applications so far have been focused on UV light detectors and electro-catalysis \cite{Mayorga-Martinez2017}.

Layered van der Waals materials coupled with 3D ferromagnets have recently been used to explore SOTs, demonstrating promising efficiencies and interesting effects \cite{Zhang2016, Shao2016, MacNeill2017, MacNeill2017b, Guimaraes2018, Stiehl2019}.
In particular, monolayers of two-dimensional semiconductors have shown large interfacially-generated SOTs \cite{Zhang2016, Shao2016, MacNeill2017b, Guimaraes2018}.
Moreover, it has been shown that layered van der Waals materials possessing low crystal symmetry can give rise to SOTs which are in principle forbidden by symmetry in standard systems, such as \PtPy{} \cite{MacNeill2017, MacNeill2017b}.
However, the microscopic mechanisms behind the generation of SOTs in van der Waals materials are still poorly understood.
It is theoretically predicted that SOTs with interfacial origins can give rise to both field-like (\tauA{}) and damping-like (\tauS{}) SOTs \cite{Amin2016,Amin2016a}.
These torques usually have forms $\tauA{} \propto \hat{m} \times \hat{y}$ and $\tauS{} \propto \hat{m} \times \left( \hat{m} \times \hat{y} \right)$, where $\hat{m}$ indicates the magnetization direction and $\hat{y}$ points in the direction perpendicular to the charge current.

\begin{figure*}[htbp]
    \centering
        \includegraphics{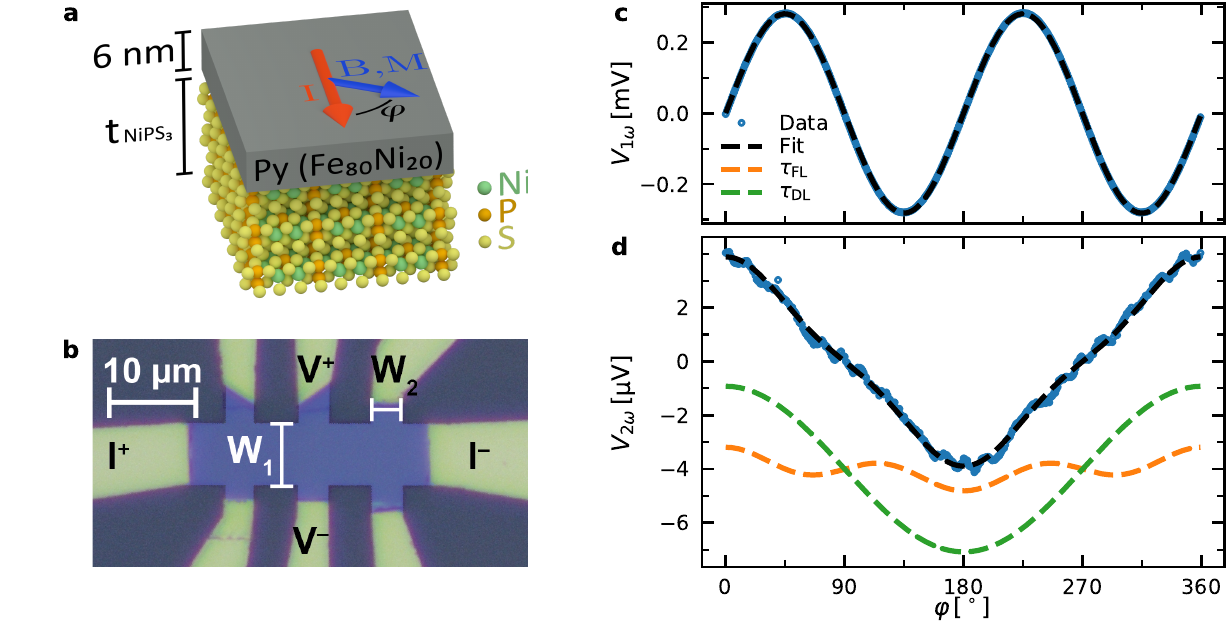}
    \caption{
        {\bf Sample geometry and typical measurements.}
        a) Schematic of the \NPSPy{} bilayers used in the second-harmonic Hall measurements.
        b) Optical micrograph of a typical Hall bar device used (Device D1).
            The light blue region in the centre is the Hall bar, patterned from \ce{NiPS3{}/Py/Al(Ox)}; the golden regions are the \ce{Au}-covered contact leads.
            Colours have been enhanced for clarity.
        First- (c) and second-(d) harmonic Hall voltage as a function in-plane magnetic field angle $\varphi$, at room temperature (\SI{300}{\kelvin}) and a magnetic field of \SI{34}{\milli\tesla} for device D1.
            The first-harmonic Hall voltage is fitted with \cref{eq:VH1} and the second-harmonic with \cref{eq:VH2}, where the individual contributions of \tauA{} and \tauS{} are separately shown, with an arbitrary offset.
            For both harmonics, a constant background has been removed.
    }
    \label{fig:Figure1}
\end{figure*}

Here we show that a \NPSPy{} bilayer device can also provide large current-induced interfacial SOTs at room temperature, with an \ipdl{} interfacial torque comparable to the best topological insulators/ferromagnet \cite{Mellnik2014} and heavy-metal (e.g.\ \ce{Pt})/ferromagnet devices \cite{Liu2011}.
In addition to the \ipdl{} torque we also observe a weaker \opfl{} torque which also is of interfacial nature.
Temperature-dependent measurements across the N\'eel temperature of \NPS{} show an increasing SOT efficiency, for both \ipdl{} and \opfl{} torques, indicating a possible influence of the magnetization ordering of \NPS{} on the SOTs \cite{Zhang2018}.
Our results demonstrate a promising route for the use of (layered) antiferromagnetic insulators for efficient manipulation of magnetic bits.

Our devices are schematically shown in \cref{fig:Figure1}a.
The device preparation is described in detail in the Methods section.
In short, thin \NPS{} crystals are mechanically exfoliated from a commercially available \NPS{} crystal (HQ Graphene).
The exfoliation is performed in vacuum, with pressures \SI{<E-6}{\milli\bar}, to maintain a high interface quality of the \NPS{} flakes.
Without breaking vacuum, \SI{6}{\nano\meter} of \Py{} is sputter-deposited on the sample followed by a thin \SI{1.5}{\nano\meter} \ce{Al} capping layer which was naturally oxidized after exposing the samples to atmosphere.
The thickness and flatness of the flakes is characterized using atomic force microscopy (AFM).
All the selected flakes for device fabrication showed a roughness below \SI{0.5}{\nano\metre} root-mean-square (RMS) in AFM images.
In the main text we focus on two devices with different values for \NPS{} layer thickness: device D1, where the \NPS{} flake has a thickness of $t_{NPS} =$ \SI{3.15}{\nano\metre}, and device D2, with $t_{NPS} =$ \SI{6.34}{\nano\metre}, corresponding to $4$ and $9$ layers of \NPS{} \cite{Kuo2016}, respectively.
Measurements for one additional device and for different current-voltage configurations can be found in the Supplementary Materials.
The Hall bars were then defined using standard electron-beam lithography and ion-beam milling techniques, followed by another lithography step and electron-beam evaporation to define the \ce{Ti(\SI{10}{\nano\metre})/Au(\SI{100}{\nano\metre})} leads.
\Cref{fig:Figure1}b shows an optical micrograph of a finished device.

\section{Results and Discussion}
The harmonic Hall measurements were performed using standard low-frequency (\SI{17}{\hertz}) lock-in techniques.
A current $I_0 \approx \SI{2.5}{\milli\ampere}$ was driven between the outer contacts and the induced Hall voltage, in the first ($V_\mathrm{H}^{1\omega}$) and second ($V_\mathrm{H}^{2\omega}$) harmonic of the frequency used, was detected between the arms of the Hall bar.
Simultaneously, a magnetic field $\bm{B}$ was applied in the sample plane under an angle $\varphi$ with respect to the current direction (\cref{fig:Figure1}a).
Assuming the magnetization $\bm{M}$ of the \Py{} layer aligns to the external magnetic field, $V_\mathrm{H}^{1\omega}$ is given by:
\begin{equation}
    \label{eq:VH1}
    V_\mathrm{H}^{1\omega} = 
        I_0 R_\mathrm{P} \sin2\varphi \sin^2\vartheta + 
        I_0 R_\mathrm{A} \cos\vartheta,
\end{equation}
\noindent where $\vartheta$ is the polar angle of the magnetic field (i.e.\ the angle with respect to the sample normal), $R_\mathrm{P}$ is the planar Hall resistance and $R_\mathrm{A}$ is the anomalous Hall resistance.

The first harmonic Hall voltage as a function of $\varphi$ for a fixed value of $B = \SI{34}{\milli\tesla}$ at room temperature (\SI{300}{\kelvin}) for device D1 is shown in \cref{fig:Figure1}c as an example of a typical measurement (other measurements on both the same device and other devices have similar signal-to-noise ratios and curve fitting quality).
The measurement is corrected for a small phase offset caused by a small misalignment of the current direction with the $x$-axis of our experimental set-up.
We observe a $\sin(2\varphi)$-behaviour with the values for $R_\mathrm{P}$ in our devices obtained by fitting our measurement using \cref{eq:VH1}.
Similarly, we extract the value for $R_\mathrm{A}$ through out-of-plane magnetic field measurements, as detailed in the Supplementary Information.
The values $R_\mathrm{P}$ and $R_\mathrm{A}$ are used to quantify the measured spin-torque values that we discuss later in the text.

Bulk \NPS{} belongs to the symmetry group $C2/m$ in its paramagnetic state \cite{Bernasconi1988,Wildes2015,Kim2019,Kuo2016}, presenting one rotation axis, a glide mirror plane, and an inversion point.
Below the N\'eel transition temperature, the magnetic texture further reduces the symmetries of the bulk to a single mirror plane, space group $Pm$ \cite{Wildes2015}.
Therefore, one could expect an induced magnetic anisotropy as reported for the low-symmetry layered materials (e.g.\ \ce{WTe2} and \ce{TaTe2}) \cite{MacNeill2017, MacNeill2017b,Stiehl2019}.
Moreover, the antiferromagnetic ordering of \NPS{} could also induce an exchange bias on the \Py{} if the magnetic structure is not strictly collinear or a small exchange bias via a perpendicular coupling at the interface of the antiferromagnetic spins of \NPS{} and the ferromagnetic spins of \Py{} \cite{Matsuyama2000}.
In the measurement shown in \cref{fig:Figure1}c we do not find a significant deviation from the fit with \cref{eq:VH1}, which does not take into account any anisotropy or exchange bias.
Hence, we do not observe an induced magnetic anisotropy or exchange bias induced by our \NPS{} crystals (see Supplementary Information for further measurements).
The lack of an induced in-plane magnetic anisotropy is in agreement with measurements in devices based on high-symmetry transition metal dichalcogenide (TMD) crystals \cite{Zhang2016, Shao2016, Guimaraes2018}.
This indicates that the magnetic anisotropy as observed in the low-symmetry materials is most likely strongly dependent on the specifics of the electronic properties and exchange coupling of the bilayer structure.
Moreover, the lack of an observable exchange bias upon field-cooling the device through the N\'eel temperature agrees with the expected collinear magnetic ordering in our \NPS{} crystals.

In addition to a first harmonic response, the presence of a non-negligible current-induced SOT gives rise to a Hall voltage in the second harmonic of the current \cite{Pi2010,Hayashi2014}.
Assuming that the magnetization direction follows the applied magnetic field, the second harmonic Hall voltage, $V_\mathrm{H}^{2\omega}$, is given by \cite{MacNeill2017b}:

\begin{equation}
    \label{eq:VH2}
    V_\mathrm{H}^{2\omega} = 
        -  I_0 R_\mathrm{P} C_\mathrm{FL}
            \cos2\varphi \cos\varphi
        \\
        -  \frac{1}{2} I_0 R_\mathrm{A} C_\mathrm{DL}
            \cos\varphi,
\end{equation}

\noindent where $C_\mathrm{FL}$ and $C_\mathrm{DL}$ are coefficients proportional to the \opfl{} and \ipdl{} torques and given by $C_\mathrm{FL} = \frac{\tauA{}}{\gamma B}$, and $C_\mathrm{DL} = \frac{\tauS{}}{\gamma \left(B+B_\mathrm{K}\right)} + \frac{2 V_\mathrm{ANE}}{I_0 R_\mathrm{A}}$.
Here $\gamma$ is the gyromagnetic ratio and $B_\mathrm{K}$ the total effective anisotropy field, including demagnetization and anisotropy terms, and $V_{ANE}$ is the anomalous Nernst contribution.

\Cref{fig:Figure1}d shows a typical measurement of the second harmonic Hall voltage as a function of the in-plane magnetic field angle $\varphi$.
The contributions by the different torques \tauA{} and \tauS{} can be obtained by fitting our data using the equation above, and their individual contributions to the fit are shown in \cref{fig:Figure1}d.
To disentangle other unwanted contributions on our signals, such as the anomalous Nernst effect \cite{Shao2016} we perform angular dependence measurements for various values of applied magnetic field \cite{Manchon2019,Avci2004}.

\begin{figure}
    \includegraphics{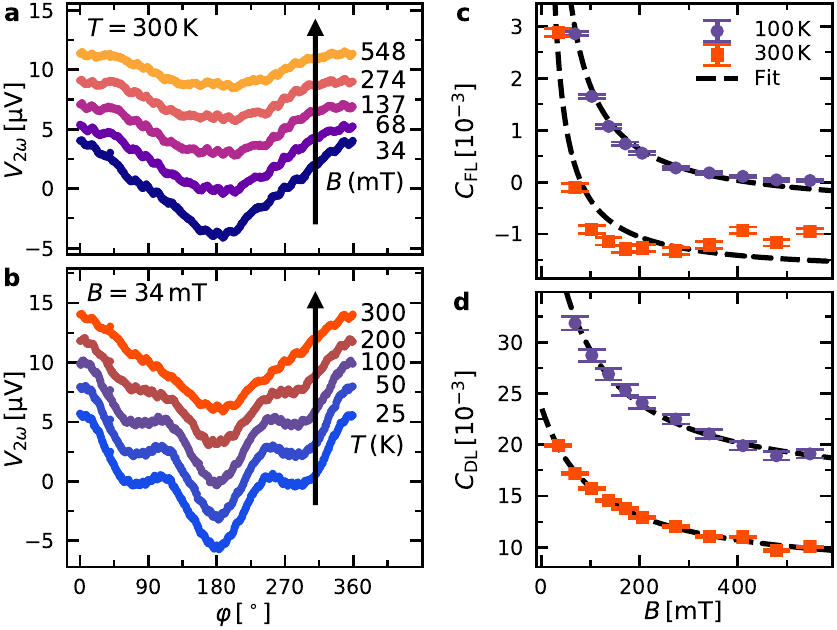}
    \caption{
        {\bf Field and temperature dependence of the second-harmonic Hall voltage.}
        The second-harmonic Hall voltage $V_{2\omega}$ is measured as a function of in-plane field angle $\varphi$ for a number of magnetic field strengths $B$ (a) and temperatures (b).
        For clarity an arbitrary offset has been added to the measurements.
        c) and d) show the magnetic field dependence of coefficients $C_\mathrm{FL}$ and $C_\mathrm{DL}$ for different temperatures.
        The fits are performed as described in the text.
    }
    \label{fig:Figure2}
\end{figure}

\Cref{fig:Figure2} shows second harmonic Hall measurements for various values of the external magnetic field strength and temperatures.
Here we see that the line shape of the second harmonic Hall measurement changes with both the external magnetic field for a fixed temperature (\cref{fig:Figure2}a), and as a function of temperature for a fixed magnetic field (\cref{fig:Figure2}b), indicating a change in weight for the different contributions for \tauS{} and \tauA{} in our devices.
We fit the angle-dependent measurements for different fields and temperatures using \cref{eq:VH2}.
There we include a constant offset to account for terms unrelated to current-induced spin-orbit torque such as thermal effects (anomalous Nernst effect) \cite{Shao2016}, and use $R_\mathrm{A}$ and $R_\mathrm{P}$ as determined earlier.
Values for the coefficients $C_\mathrm{FL}$ and $C_\mathrm{DL}$ are obtained for each individual measurement, shown in \cref{fig:Figure2}c and d.
The current-induced SOTs are then quantified by fitting $C_\mathrm{FL}$ and $C_\mathrm{DL}$, as described earlier, shown as dashed lines in \cref{fig:Figure2}c and d, to extract \tauA{} and \tauS{}.
It has been reported that spin-orbit torque measurements using the second harmonic Hall technique can be influenced by the aspect ratio of the Hall bar dimensions ($W_2/W_1$ as specified in \cref{fig:Figure1}a) \cite{Neumann2018}.
Therefore, in order to better quantify our results, the values for the torques (i.e.\ \tauA{} and \tauS{}) we obtained were corrected for our specific Hall bar geometry by dividing the torque value by a factor corresponding to the Hall bar geometry \footnote{The SOT values for D1 were corrected by dividing the obtained values by a factor of $0.67$ (for $W_2/W_1 = 0.83$) and for D2 by a factor of $0.72$ (for $W_2/W_1 = 0.67$) \cite{Neumann2018}.}.

At room temperature (\SI{300}{\kelvin}) we observe an \ipdl{} [$\hat{m} \times \left( \hat{m} \times \hat{y} \right)$] torque $\tauS{}/(\gamma I_0) = \SI{1.0(1)}{\milli\tesla\per\milli\ampere}$, for device D1.
For a better comparison between our devices and others in literature, the spin orbit torque can be normalized by the electric field ($E$) applied to the device, $\tauS{}/(\gamma E)$. The torque value can also be evaluated as torque conductivity $\sigma$, defined as the angular momentum absorbed by the magnet per second per unit interface area per unit electric field.
For a torque $\tau_i$ ($i$= DL or FL), we calculate the corresponding spin-torque conductivity by $\sigma = M_\mathrm{S} t_\mathrm{FM} \frac{W_1}{R_\mathrm{S}} \frac{\tau_i}{\gamma I_0}$,
where $M_\mathrm{S}$ is the saturation magnetization of the \ce{Py} FM layer, $t_\mathrm{FM} = \SI{6}{\nano\meter}$ is the thickness of the \ce{Py} layer, $W_1$ is the Hall bar width as defined in \cref{fig:Figure1}a, and $R_\mathrm{S}$ is the sheet resistance of the measured device; torque conductivities for various devices are summarized in \cref{tab:torque_conduct_table}.
We obtain $\tauS{}/(\gamma E) = \SI{22(3)}{\nano\meter\tesla\per\volt}$ and  $\sigma_{DL} = \SI{2.2(3)e5}{(\frac{\hbar}{2e}) \per(\ohm\meter)}$ for the \ipdl{} torque \tauS{}, using $\mu_0 M_\mathrm{S} = \SI{0.7}{\tesla}$ as discussed in the Supplementary Information \cite{MacNeill2017,MacNeill2017b,Guimaraes2018,Stiehl2019}.
This is the largest damping-like torque conductivity for all layered material/ferromagnet devices reported so far, which are over one order of magnitude lower \cite{Zhang2016, Shao2016, MacNeill2017, MacNeill2017b, Guimaraes2018, Stiehl2019,Stiehl2019a}, and is in the range of the best values obtained using standard heavy-metal/ferromagnet devices and topological insulator/ferromagnet devices, in the order of \SI{1e5}{(\frac{\hbar}{2e}) \per(\ohm\meter)} \cite{Nguyen2016, Mellnik2014, Manchon2019}.

\begin{table*}
    \centering
    \caption{
        The torque conductivities \sigmaA{} and \sigmaS{} [in \si{10^{5} (\frac{\hbar}{2e}) \per(\ohm\meter)}] is given at both room temperature (RT, \SI{300}{\kelvin}) and low temperature (LT, \SI{50}{\kelvin}), assuming a saturation magnetization of $\mu_0 M_\mathrm{S} = \SI{0.7}{\tesla}$.
        Also the dimensions of the Hall bars ($W_1$ and $W_2$ in \si{\micro\metre} as defined in \cref{fig:Figure1}), the thickness of the \NPS{} flake $t_\mathrm{NPS}$ (in \si{\nano\metre}), with the estimated number of layers $n$ \cite{Kuo2016} between brackets, and the sheet resistance $R_\mathrm{S}^\mathrm{RT}$ (in \si{\ohm\per\sq}) at room temperature are given.
    }
    \label{tab:torque_conduct_table}
    \begin{tabular}{
                c
                S[table-format = 1.1(0)]
                S[table-format = 1.1(0)]
                c
                S[table-format = 3.1(0)]
                S[table-format = 1.4(1)]
                S[table-format = 2.3(1)]
                S[table-format = 2.3(1)]
                S[table-format = 2.1(1)]
            }
        \toprule
        {Device} & {$W_1$} & {$W_2$} & {$t_\mathrm{NPS}$ ($n$)} & {$R_\mathrm{S}^\mathrm{RT}$} &
            {$\sigma_\mathrm{FL}^\mathrm{RT}$} & {$\sigma_\mathrm{FL}^\mathrm{LT}$} &
            {$\sigma_\mathrm{DL}^\mathrm{RT}$} & {$\sigma_\mathrm{DL}^\mathrm{LT}$} \\
        \midrule
              D1 &   3.0 &    2.5 &   3.1 (4) &  140.7 &    0.17(2) &   0.319(4) &     2.2(3) &     3.2(3) \\
              D2 &   3.0 &    2.0 &   6.3 (7) &  151.8 &     0.1(1) &   -2.32(8) &     0.6(1) &      10(2) \\
              D3 &   3.0 &    2.0 &   5.2 (9) &  144.3 &       {--} &   -2.01(8) &       {--} &       8(1) \\
         \PtPy{} &   5.0 &    3.0 &      {--} &   15.1 &    7.76(3) &    8.99(3) &     7.3(3) &       8(1) \\
           \Py{} &   5.0 &    3.0 &      {--} &  112.0 &  0.0921(9) &       {--} &  -0.079(9) &       {--} \\
        \bottomrule
    \end{tabular}
\end{table*}

We also find a non-negligible \opfl{} torque ($\hat{m} \times \hat{y}$) in our \NPSPy{} devices, $\tauA{}/(\gamma I_0) = $\SI{0.079(9)}{\milli\tesla\per\milli\ampere} at room temperature, corresponding to $\tauA{}/(\gamma E) = \SI{1.7(2)}{\nano\meter\tesla\per\volt}$ and a spin-torque conductivity of $\sigma_{FL} = $\SI{1.7(2)e4}{(\frac{\hbar}{2e}) \per(\ohm\meter)}.
The presence of both damping-like and field-like torques arising from interfacial SOTs are in agreement with theoretical predictions \cite{Amin2016,Amin2016a}.
However, the magnitude for \tauA{} is about one order of magnitude smaller than the damping-like torques discussed above.

To understand where the observed torque originates from it is instructive to consider the possible current paths through the \NPSPy{} bilayer.
Intrinsic \NPS{} is highly resistive \cite{Foot1980,Chu2017}, in sharp contrast to the metallic \Py{} layer, which has a resistivity that is orders of magnitude lower than \NPS{} (of the order of \SI{e-5}{\ohm\centi\meter} for \Py{} compared to \SI{e11}{\ohm\centi\meter} for \NPS{}).
Hence, we expect that all current flows through the \Py{} layer in our \NPSPy{} devices.
This is confirmed by measurements of the sheet resistances $R_\mathrm{S}$ for \NPSPy{} based devices (\num{140} to \SI{150}{\ohm\per\sq}) and a \Py{} based device (\SI{\sim 110}{\ohm\per\sq}), i.e. without a SOT material layer.
The small difference in sheet resistance could be attributed to a difference in quality of the \Py{} layer when grown on top of the different surfaces, the \NPS{} flake or the bare \ce{Si/SiO2} substrate.
Hence, the torques measured have to arise from the interface between \Py{} and \NPS{}; the large values for \tauS{} observed in our devices indicate the presence of a very strong interfacial SOT.

Other possible contributions to the observed SOTs can arise from the \ce{Al} capping layer. 
If the \ce{Al} capping layer is not completely oxidized, a current path through the \ce{Al} capping layer allows the generation of an Oersted field working on the \Py{} layer.
Alternatively, when a current flows through the \Py{} layer in an inhomogeneous manner, the Oersted field generated by the current through the \Py{} layer do not fully cancel and a net \opfl{} torque can be measured.
In order to probe such possible contributions on our results we perform control measurements in devices based on a single \Py{} layer, without a SOT material, but still capped with the naturally-oxidized \ce{Al(\SI{1.5}{\nano\metre})} layer.
For these samples we obtained $\tauS{}/(\gamma E) = \SI{-0.78(9)}{\nano\meter\tesla\per\volt}$
, considerably smaller (and of opposite sign) than the values obtained in our \NPS{} devices.
Interestingly, we also observe a measurable, albeit smaller, \tauA{} for Hall bars based on only \Py{} $\tauA{}/(\gamma E) =  \SI{0.91(9)}{\nano\meter\tesla\per\volt}$
, i.e.\ without a spin-orbit torque generating material.
This unexpected torque could be an indication an additional contribution from either an unoxidised portion of the \ce{Al} capping layer or an inhomogeneous current distribution in the ferromagnetic layer.
However, as the SOTs observed in this device are significantly smaller than the SOTs observed in the \NPS{} devices, this shows that the Al oxide capping layer has a minimal effect in our measured torque values and points to the crucial role of the \NPS{} flake on the measured spin-orbit torques in those devices.

We also perform a direct comparison to standard heavy-metal/ferromagnet SOT devices by performing control measurements in a \PtPy{} device fabricated using the same procedure as the \NPS{} devices.
For these devices we obtain $\tauS{}/(\gamma E) =  \SI{72(3)}{\nano\meter\tesla\per\volt}$.
This translates to a torque conductivity $\sigmaS{} = \SI{7.3(3)e5}{(\frac{\hbar}{2e}) \per(\ohm\meter)}$, which is in line with the typical torques observed in literature \cite{Nguyen2016, Mellnik2014, Manchon2019}.
This torque is only slightly larger than the torque found in the \NPS{} based device, illustrating that the torque found in the \NPS{} based device is indeed in the range of the best heavy-metal based or topological insulator based devices.
Finally, for our \PtPy{} devices we observe a $\tauA{}/(\gamma E) =  \SI{76.4(3)}{\nano\meter\tesla\per\volt}$
, with a magnitude consistent with the expected Oersted-field contribution from the current flowing in the \ce{Pt} layer.

Even though a significant \opfl{} torque is observed in all our devices, there is a clear difference between the results in our control \Py{} and \PtPy{} devices and the ones for the \NPSPy{} devices: while \Py{} and \PtPy{} devices consistently show a positive sign for \tauA{}, we observe both a positive and negative signs for our \NPSPy{} devices (see Supplementary Information for more measurements).
The presence of an interfacial \opfl{} torque has been observed in devices based on TMD monolayers \cite{Shao2016}, with a sign change with respect to Oersted-fields observed in monolayer \ce{NbSe2{}/Py} bilayers \cite{Guimaraes2018}.
Albeit we cannot completely rule out the possibility of alternative mechanisms for the observed \tauA{}, such as an inhomogeneous current distribution in the \Py{} layer, the comparison of the results for the three different devices (\Py{}, \PtPy{}, and \NPSPy{}) indicate the presence of a non-negligible interfacial \opfl{} torque.

\begin{figure}
    \includegraphics{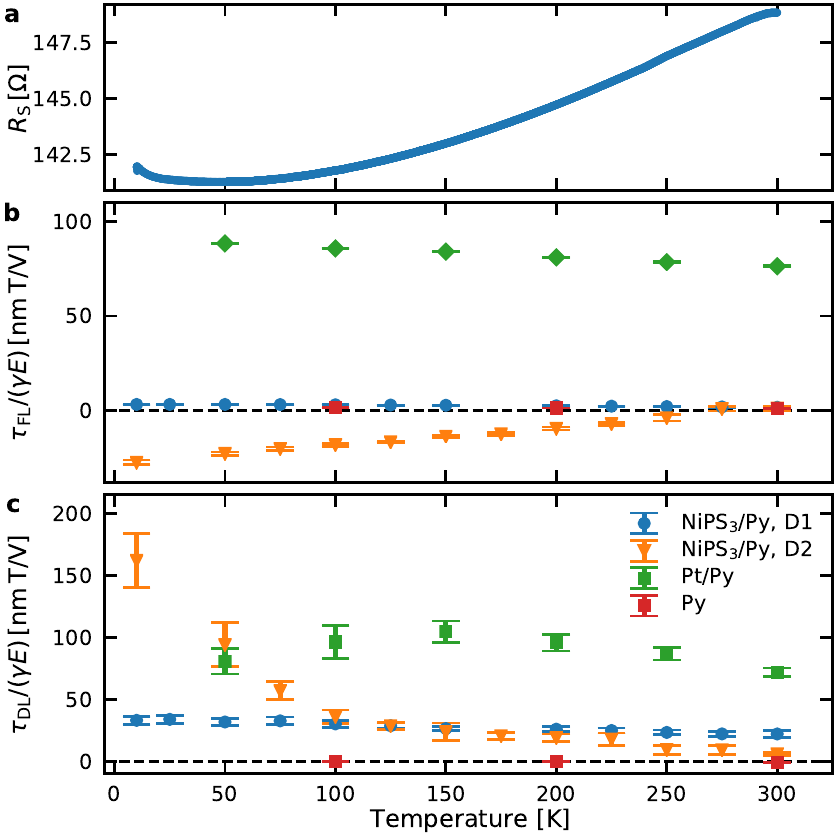}
    \caption{
        {\bf Torques as a function of temperature.}
        a) Sheet resistance $R_\mathrm{S}$ of the \ce{NiPS3{}/Py/Al(Ox)} Hall bar as a function of temperature.
        Out-of-plane field-like torque \tauA{} (b) and \ipdl{} torque \tauS{} (c) as a function of temperature in the \NPSPy{} bilayer devices, a \Py{} reference sample, and a \PtPy{} reference sample.
        The measured torques have been corrected for influence of the Hall bar geometry, according to \cite{Neumann2018}, and normalized by the applied electric field for better comparison among devices.
    }
    \label{fig:Figure3}
\end{figure}

In order to explore the effect of the antiferromagnetic phase transition of \NPS{} on the current-induced SOTs we performed measurements as a function of temperature across $T_\mathrm{N} \sim \SI{170}{\kelvin}$, with $T$ ranging from \SIrange{10}{300}{\kelvin}.
Our devices show a decrease in sheet resistance of about 5$\%$ with a decrease in temperature indicating a metallic behaviour (\cref{fig:Figure3}a for device D1), in agreement with the expectation that the resistance in our devices is dominated by the metallic \Py{} layer.

By performing the same procedure for quantifying \tauS{} and \tauA{} described above, we extract the temperature dependence of the spin-orbit torques (\cref{fig:Figure3}b and c for \tauS{} and \tauA{} respectively).
In \cref{tab:torque_conduct_table} the torque conductivities at room temperature and low temperature are summarized for all devices.
We observe that there is indeed a temperature dependence of both the \opfl{} and \ipdl{} torques; in device D1 the torques increase up to  \SI{3.16(3)}{\nano\meter\tesla\per\volt} and \SI{33(3)}{\nano\meter\tesla\per\volt}
when lowering the temperature to \SI{10}{\kelvin} for the \opfl{} and \ipdl{} torques, respectively.
For the \ipdl{} torque this behaviour is consistent over different devices: the torque value increases when the temperature decreases, though the amount of increase is not constant over the different devices (see Supplementary Information).
We find no consistent behaviour for the \opfl{} torque in our \NPSPy{} devices apart from the fact that the torque magnitude and sign seem to be strongly dependent on temperature (see Supplementary Information).

The increase for the \ipdl{} torque with a decrease in temperature seems reproducible among our \NPSPy{} devices, however, we find that the specific trend with temperature is device specific.
For device D2 for example, we observe a smaller \ipdl{} torque at room temperature ($\tauS{}/(\gamma E) =  \SI{6(1)}{\nano\meter\tesla\per\volt}$
) and a much steeper increase to  \SI{160(20)}{\nano\meter\tesla\per\volt}
at \SI{10}{\kelvin}, significantly larger than the maximum value in device D1.
Interestingly, for all our devices we see an onset on the change in torque magnitude at temperatures near the N\'eel temperature of \NPS{}, around \SI{150}{\kelvin}.
Even though this is a qualitative observation, we believe that this is an indication that the magnetic ordering in \NPS{} has an effect on the measured SOTs.
However, further research is required to confirm whether or not the magnetic order of \NPS{} is responsible for (part of) the observed temperature dependence.

The behaviour of the \opfl{} torque \tauA{} is different for different devices.
For some devices we observe a negative value for \tauA{} that also increases in magnitude when the temperature is decreased while others show an initially positive torque that changes sign when the temperature is decreased.
The reason for these different temperature dependencies (for both the \ipdl{} and \opfl{} torque) remains unclear and requires further studies.
Possible explanations might be related to the thickness of the \NPS{} flake -- device D1 contains a relatively thin flake of only $4$ layers of \NPS{} while device D2 contains a flake of $9$ layers -- or to the quality of the interface between the \NPS{} flake and the \Py{} layer, which could e.g.\ result in a temperature-dependent spin-mixing conductance.
Although devices with 3 different \NPS{} thicknesses were measured and both similarities and differences were found, we did not observe a systematic behaviour with thickness.
A more systematic study on the thickness dependence is required and is left for future investigations.

We now compare the temperature dependence of the SOTs obtained for our \NPSPy{} to our control \PtPy{} and \Py{} devices, \cref{fig:Figure3}b and c.
For the \Py{} based device we observe a very small change in both \tauA{} and \tauS{}, with values close to zero throughout the whole measured temperature range.
For the \PtPy{} device we find a small monotonic increase of \tauA{} with a decrease in temperature, probably indicating that the Pt layer decreases its resistivity faster than the Py layer, therefore increasing the Oersted-field torque in this device.
We also observe a small change in \tauS{} with a change in temperature, but different from the monotonic increase observed for our \NPSPy{} devices.
This strengthens the conclusion that both the SOTs observed in the \NPSPy{} devices and their temperature dependence originate from the \NPSPy{} interface.

While the exact origin of the observed spin-orbit torque in the \NPSPy{} devices is not understood at this moment, we suggest two possible mechanisms.  
First, the inversion symmetry breaking at the \NPSPy{} interface can allow for the presence of a Rashba spin-orbit coupling.
This effect has been shown to give rise to strong SOTs in metallic \cite{Manchon2015}, topological-insulators \cite{Mellnik2014,Manchon2015} and TMD-based devices \cite{MacNeill2017,MacNeill2017b}. 
Alternatively, a non-collinearity of the antiferromagnetic order in \NPS{} or the ferromagnetic ordering in \Py{} could allow for an effective SOTs to generated even in the absence of a spin-orbit interaction in the \NPS{} \cite{Zhang2018}.
Though both \NPS{} and \Py{} present fully collinear magnetic structures, their mutual exchange interaction can lead to a spatially-varying magnetic ordering, thereby adding a small non-collinear contribution to the magnetic order.
However, a more in-depth understanding of the nature of the mechanisms involved in generating the observed SOTs requires a more thorough theoretical treatment.

We point out that intermixing or a magnetic dead layer at the interface between \Py{} and NiPS3 can also affect the torques.
It has recently been demonstrated that a single \Py{} layer can also generate SOTs in an asymmetric stack \cite{Wang2019b}.
These anomalous spin-orbit torques are strongly dependent on the properties of both interfaces and at this moment their contribution in previously measured spin-orbit torques is still unclear.
However, the net torques estimated for insulator/thin \Py{}/insulator structures were still one order of magnitude smaller than the torques measured here.
A careful thickness dependence combined with cross-sectional transmission electron microscopy could help to clarify the importance of such effects in future works.

Below the N\'eel temperature, \NPS{} presents an antiferromagnetic ordering (with ferromagnetic zigzag lines that couple antiferromagnetically) \cite{Wildes2015} which breaks the glide mirror plane and the screw symmetry axis.
It has been shown that a crystallographic (or magnetic) symmetry breaking can lead to non-standard spin-orbit torques \cite{MacNeill2017,MacNeill2017b, Stiehl2019,Stiehl2019a,Zhang2016a,Zhang2018}.
In order to explore a possible effect of the crystal and magnetic symmetries and orientation on the measured SOTs, we perform the fitting procedure with extra terms representing an out-of-plane damping-like torque ($\hat{m} \times \hat{m} \times \hat{z}$).
Additionally, we performed the same measurements and analysis with the current and voltage paths interchanged, i.e.\ rotated by $\pi/2$.
Here the angle between the current and the zigzag line of the magnetic ordering should change for the two configurations which could have an influence on the torques that are generated by the \NPSPy{} interface.
We observe only a small difference of a factor of approximately $1.5$ in \tauS{} and smaller for \tauA{} (see Supplementary Information).
As this is reproduced in our control devices it likely arises from the different current paths for the two configurations and seems to be unrelated to the crystal properties of \NPS{}.
Therefore, no torque components related to the crystal and magnetic symmetries and orientation are observed within our experimental accuracy.

\section{Conclusion}
In conclusion, we observe large interfacial \ipdl{} SOTs in \NPSPy{} bilayers.
Our devices present \ipdl{} SOTs in the order of $\tauS{}/(\gamma E) =$ \SI{20}{\nano\meter\tesla\per\volt} and \SI{80}{\nano\meter\tesla\per\volt} at room and low temperatures, respectively, compared to \SI{70}{\nano\meter\tesla\per\volt} found a heavy-metal/ferromagnet device (\PtPy{}).
Additionally, we observe a small interfacial \opfl{} SOT of $\tauA{}/(\gamma E) =$ \SI{2}{\nano\meter\tesla\per\volt}
with a direction that is opposite to a torque from Oersted-fields coming from a current through the \NPS{} flake, which is in line with the high resistivity of \NPS{} that prevents a current from running through the flake.
Though we observed a (non-trivial) temperature dependence of the observed SOTs, we found no clear relation to the antiferromagnetic phase-transition of \NPS{} or the related reduction of the crystallographic symmetry.
Based on these findings, we conclude that there is a significant contribution of the interface between \NPS{} and \Py{} to both the \opfl{} and \ipdl{} SOTs, although the microscopic origin is not yet understood.

Our results add to the understanding that the detailed electronic structure of the interface between the spin-orbit material and ferromagnet plays a critical role on the measured SOTs, and should encourage the development of a more complete theoretical framework for the prediction of SOTs using various materials.
The large interfacial torque and lack of dependence on the specific crystal symmetries or orientation is ideal for highly-efficient SOT devices.
The fact that current flows only through the ferromagnetic layer allows for the use of lower total currents for magnetization switching when compared to standard heavy-metal/ferromagnet devices.
Therefore, we believe our results illustrate the potential of insulating van der Waals crystals for spintronic applications.

\section{Acknowledgements}
We acknowledge B. Koopmans and M. Titov for fruitful discussions and for useful comments on the manuscript, and J. Francke and G. Basselmans for technical help with the experimental setup.
Sample fabrication was performed using NanoLabNL facilities.
The research performed here was funded by the Dutch Research Council (NWO) under the grants VENI 15093 and 680-91-113.

\section{Methods}
\subsection{Sample fabrication}
\NPS{} flakes were mechanically exfoliated from commercially available crystals (HQ Graphene) onto a thermally oxidized \ce{Si/SiO2} substrate (with \SI{100}{\nano\metre} \ce{SiO2}).
For this exfoliation ordinary scotch tape was used.
To prevent degradation of the flakes and conserve the high-quality interface of the exfoliated flakes the exfoliation is performed in two steps.
First the tape with flakes is prepared and placed on the substrate in a nitrogen-filled glove-box.
The substrate, with tape, is transported (through air) to the load-lock of the deposition system.
Here, the actual exfoliation is performed when the load-lock has reached a pressure \SI{<e-6}{\milli\bar}.

The sample is then immediately transported, through vacuum, into the deposition chamber, where \ce{Py(6)/Al(1.5)} is deposited on the sample by magnetron sputtering.
Afterwards, the sample is taken out of the vacuum into air, where the \ce{Al} layer will oxidise to \ce{Al2O3}, creating an insulating and protective layer for the sample.
Using an optical microscope the sample is inspected to find sufficiently large (i.e.\ larger than \SI{5x5}{\micro\metre}) \NPS{} flakes for later sample fabrication.

For each sufficiently large flake a Hall bar is designed and patterned into a \ce{SiO2} hard-mask using electron-beam lithography (EBL) with poly-(methyl-methacrylate) (PMMA), sputter deposition of \ce{SiO2 (\SI{60}{\nano\metre})}, and a lift-off process.
The Hall bar is then etched into the \NPS{} flake using argon (\ce{Ar}) ion-beam milling, a layer of \ce{SiO2 (\SI{20}{\nano\metre})} is sputter deposited to clamp the Hall bars on the sample.

Hereafter the Hall bars contacts are fabricated.
The contacts are patterned into a layer of PMMA, again using EBL.
Using reactive ion etching the \ce{SiO2} layer is removed in places where the contacts will be deposited.
Finally, \ce{Ti/Au} is deposited for the contacts, after a short argon etching to remove the \ce{Al2O3} on top of the \Py{} for better contacts, and lift-off is performed to remove the PMMA and the redundant \ce{Ti/Au}.


\section{Author contributions}
CFS and MHDG conceived the experiment, fabricated the samples, and performed the measurements while advised by HJMS.
CFS performed the data analysis under MHDG and HJMS supervision.
CFS and MHDG wrote the manuscript with comments from HJMS.

\bibliography{Bibliography}

\end{document}


\title{
    Supplementary material for
    \texorpdfstring{\\}{ }
    Large interfacial spin-orbit torques in
    \texorpdfstring{\\}{ }
    layered antiferromagnetic insulator \NPS{}/ferromagnet bilayers
}
\date{\today}

\author{C. F. Schippers}
\email{c.f.schippers@tue.nl}
\affiliation{Department of Applied Physics, Eindhoven University of Technology, P.O. Box 513, 5600 MB, Eindhoven, the Netherlands}

\author{H. J. M. Swagten}
\affiliation{Department of Applied Physics, Eindhoven University of Technology, P.O. Box 513, 5600 MB, Eindhoven, the Netherlands}

\author{M. H. D. Guimarães}
\email{m.h.guimaraes@rug.nl}
\affiliation{Department of Applied Physics, Eindhoven University of Technology, P.O. Box 513, 5600 MB, Eindhoven, the Netherlands}
\affiliation{Zernike Institute for Advanced Materials, University of Groningen, P.O. Box 221, 9747 AG, Groningen, the Netherlands}

\maketitle

\section{Determination of anomalous Hall resistance}
\label{sec:RAHE}
In order to obtain precise values for the spin-orbit torques (SOTs) from the second-harmonic Hall measurements, the anomalous Hall resistance needs to be determined for the measured devices.
For this a probing AC current is driven in the Hall bar and the generated transverse voltage is measured -- specifically the first harmonic transverse voltage -- while rotating the sample in an external magnetic field.
In contrast to the measurement in the main text, the sample is rotated around on of its in-plane directions such that the magnetic field (from the perspective of the sample) is rotated through the out-of-plane direction; i.e., in terms of Eq. 1 of the main text, $\theta$ is varied while $\phi$ is kept constant.

\begin{figure}[b]
    \includegraphics{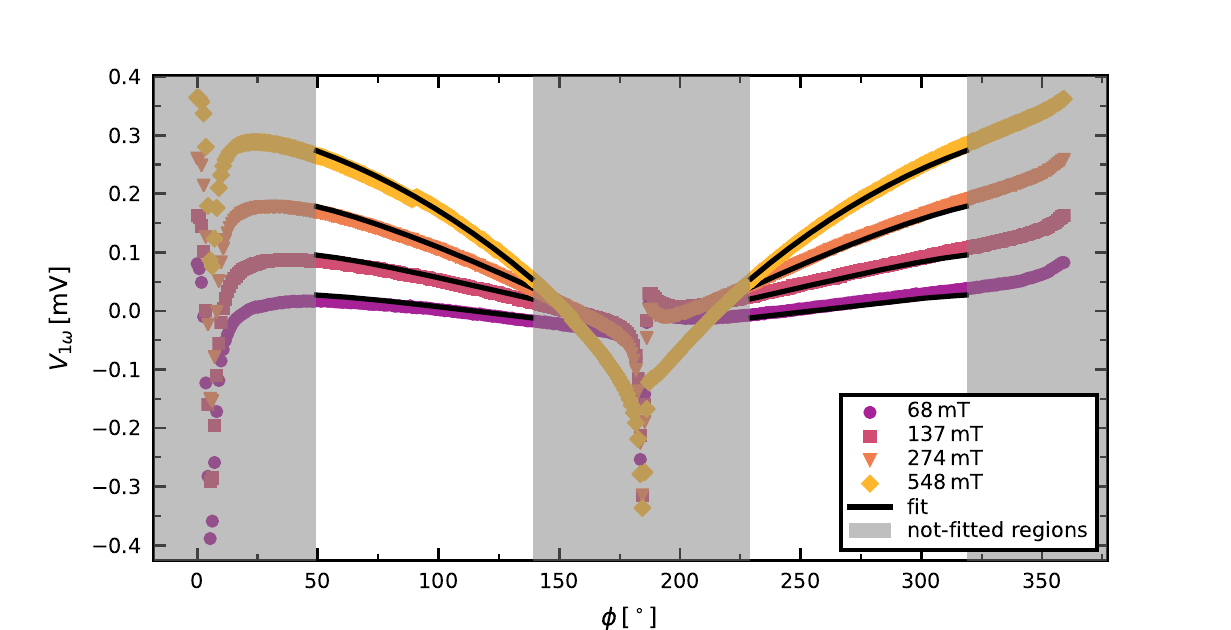}
    \caption{
        Hall voltages of a Hall bar with \ce{NiPS3{}/Py(\SI{6}{\nano\metre})} as a function of out-of-plane field-angle $\theta$ for varying magnetic field strengths at \SI{300}{\kelvin}.
        The black lines are fits using Eq. 1 of the main text, with corrections for the misalignment of the magnetic field and magnetization due to demagnetization.
    }
    \label{fig:AHE_measurement}
\end{figure}

In \cref{fig:AHE_measurement} a series of anomalous Hall measurements for different external magnetic fields is shown.
While a simple sinusoidal measurement as a function of out-of-plane field-angle $\theta$ is expected from Eq. 1 of the main text, the measurements shows a more complex behaviour.
We attribute this to a sufficiently strong in-plane shape anisotropy of the \Py{} layer due to its small thickness (\SI{6}{\nano\metre}) with respect to the other dimensions.

To account for this and extract the anomalous Hall resistance $R_\mathrm{A}$ two additional steps are added to the data analysis procedure.
First the measurement is fitted only for angles sufficiently far away from the out-of-plane direction (i.e.\ $\left|\theta\right| <= \SI{45}{\degree}$) as around the out-of-plane direction the measured Hall voltage is most influenced by the in-plane shape anisotropy.
Further a simple energy model is used to correct for a discrepancy between the out-of-plane field-angle $\theta$ and the actual out-of-plane magnetization-angle $\theta_M$; the model for calculating energy $U$ is given by:
\begin{equation}
    \label{eq:SW_model}
    U = - \mathbf{B} \cdot \mathbf{M}
        + N_z \frac{1}{2} \mu_0 m_z^2
        + N_x \frac{1}{2} \mu_0 m_x^2,
\end{equation}

where $\mathbf{B}$ and $\mathbf{M}$ are the external magnetic field and the \Py{} magnetization, respectively, $m_{(x,z)}$ is $x$ or $z$ magnetization component, $\mu_0$ is the magnetic permeability and $N_z$ and $N_x$ are the out-of-plane and in-plane demagnetization factor, respectively.
As we are considering a thin magnetic film, we assume that $N_z \approx 1$ and that $N_x \ll 0.1$; i.e.\ an overall in-plane shape anisotropy with a small anisotropy within the plane.
Moreover, we assume that the demagnetization is the dominant form of anisotropy in our system.
Finally, also a possible exchange bias from the antiferromagnetic \NPS{} is omitted since we experimentally found no evidence for an exchange bias within this system, as discussed later in \cref{sec:no_eb_an}.
In order to correct for the misalignment of the external magnetic field and the magnetization direction, a minimization routine for energy $U$ has been incorporated into the fitting procedure, resulting in the fitted curves in \cref{fig:AHE_measurement}.

Using these two corrections we are able to extract an anomalous Hall resistance from the AHE measurements.

\subsection{Additional measurements of the anomalous Hall resistance}
To show that the values for $R_\mathrm{A}$ are reasonable we measured the anomalous Hall resistance in a set-up where the magnetic field is swept between \SI{\pm2}{\tesla}, which suffices to saturate the \Py{} layer.
As using a device for this measurement rendered the sample useless for the second harmonic Hall measurements it is only performed on the \ce{Py/AlOx} device for which measurements were shown in the main text.

\begin{figure}
    \includegraphics{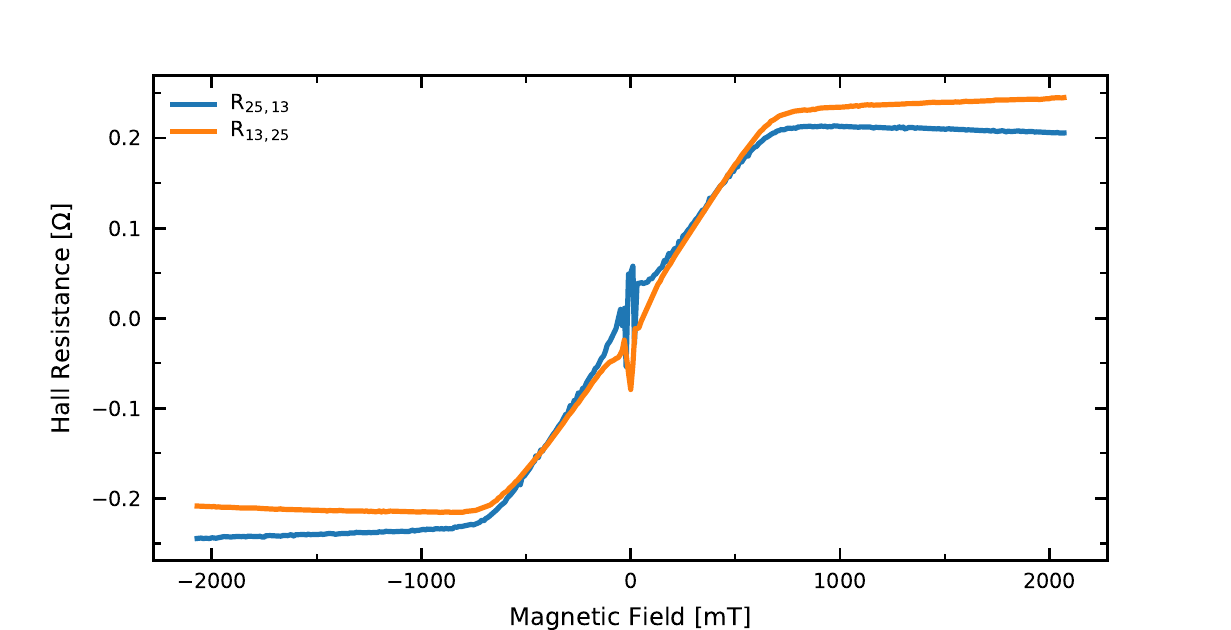}
    \caption{
        Anomalous Hall resistance of a Hall bar with \SI{6}{\nano\metre} of \Py{} as a function of out-of-plane external magnetic field.
        A linear background has been removed from the measurements.
    }
    \label{fig:AHE_measurement_Py}
\end{figure}

In \cref{fig:AHE_measurement_Py} the AHE measurement is shown for both current directions.
The resistance was measured by driving an AC current of \SI{1}{\milli\ampere} (root-mean-square) at a frequency of \SI{779}{\hertz} and measuring the transverse voltage using a lock-in amplifier.
A linear background and constant offset has been removed from the measurements to better determine $R_\mathrm{A}$.
In the measurement, we see that the resistance increases up to a field of \SI{\pm 750}{\milli\tesla}, after which the value saturates.

By averaging the (absolute value of) both high magnetic field directions we obtain $R_\mathrm{A} = \SI{0.224(4)}{\ohm}$ for the $R_{25,13}$ configuration and $R_\mathrm{A} = \SI{0.226(1)}{\ohm}$ for the $R_{13,25}$ configuration.
Similar values (up to a factor of 2) were found in similar devices using the method described in the previous section.
$R_\mathrm{A}$ is fairly similar for both current configurations, showing that we can safely use a single $R_\mathrm{A}$ value for correcting both current configurations of the second harmonic Hall measurement.
This illustrates that the used $R_\mathrm{A}$ for interpreting the second harmonic Hall measurements should not affect the drawn conclusions.

\section{Magnetometry measurement of \Py{} layer}

\begin{figure}
    \includegraphics{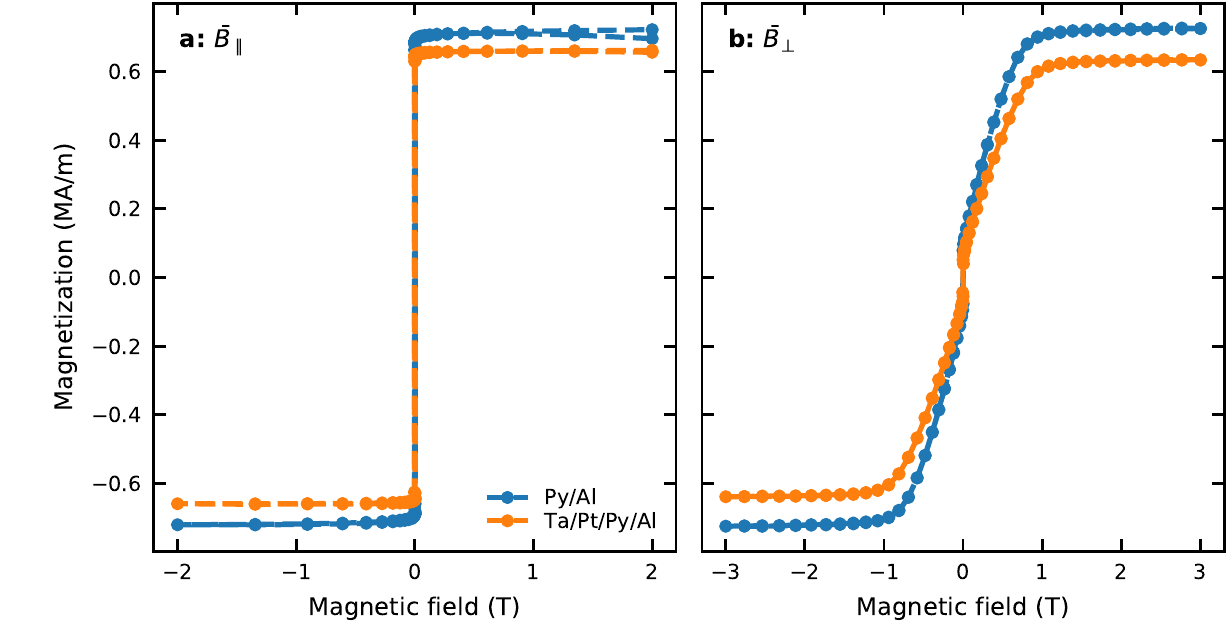}
    \caption{
        Magnetometry measurement of a \ce{Py(6)/Al(1.5)} stack (blue) and a \ce{Ta(3)/Pt(3)/Py(6)/Al(1.5)} stack (orange) using a VSM-SQUID with the field parallel to the sample plane (a) and perpendicular to the sample plane (b).
        The measurements have been corrected for shape-artifacts and a linear background is subtracted.
    }
    \label{fig:VSM_Py}
\end{figure}

Here we investigate the saturation magnetization of our \Py{} thin films, required for estimating the torque conductivity $\sigma$ in the main text.
To this end, full-sheet films of \ce{Py(6)/Al(1.5)} and \ce{Ta(3)/Pt(3)/Py(6)/Al(1.5)} were grown and measured using a VSM-SQUID.
The measurements are shown in \cref{fig:VSM_Py}, both for the magnetic field parallel to the sample plane and the magnetic field perpendicular to the sample plane.
In both configurations we observe that the magnetization of \ce{Py/Al} saturates to \SI{0.72}{\mega\ampere\per\metre}, equivalent to $\mu_0 M_\mathrm{S} = \SI{0.90}{\tesla}$.
The \ce{Ta/Pt/Py/Al} sample gives a saturation magnetization of \SI{0.65}{\mega\ampere\per\metre}, or $\mu_0 M_\mathrm{S} = \SI{0.82}{\tesla}$.

Device fabrication, which includes heating and ion bombardment of the sample, can significantly influence the magnetization of the \Py{} layers in the devices used for the measurements in the main text.
For the \ce{Py/Al} device, the saturation magnetization can be estimated from the measurement presented in \cref{fig:AHE_measurement_Py}.
Assuming the anisotropy of the \Py{} layer consists only of shape anisotropy, the field at which the Hall resistance saturates, and hence the field at which the magnetization is completely saturated in the out-of-plane direction, is equal to the saturation magnetization of the \Py{} layer.
From this, we estimate the saturation magnetization in this device to be $\mu_0 M_\mathrm{S} = \SI{0.7}{\tesla}$.
This is slightly smaller than the saturation magnetization observed in the full-sheet films, and this difference can indeed be attributed to the device fabrication.
Therefore, we adopt a saturation magnetization of \SI{0.7}{\tesla}, as obtained in our devices, for the calculation of the torque conductivities.

\section{Effect of the current direction on the SOTs}
\begin{figure}
    \includegraphics{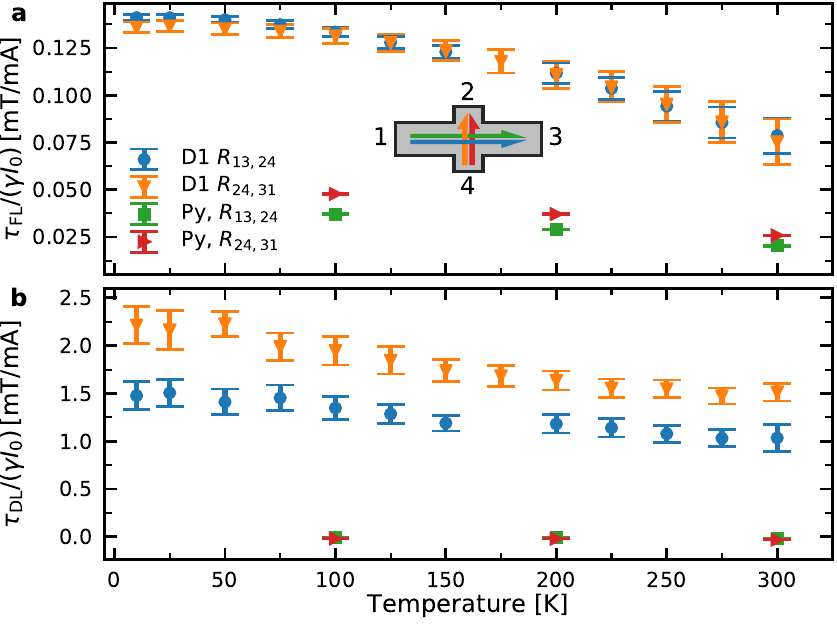}
    \caption{
        Measured SOTs as a function of temperature temperatures for the \opfl{} torque \tauA{} (a) and the \ipdl{} torque \tauS{} (b) in the \NPSPy{} device and a \Py{} reference sample.
        The torques have been measured for two difference current directions as indicated in the sketch in (a); the voltage is always measured perpendicular to the applied current.
        $R_{ab,cd}$ indicates that the current is driven between contacts $a$ and $b$ and the Hall voltage is measured between contacts $c$ and $d$.
        The torques have been corrected for the Hall bar geometry according to \cite{Neumann2018}.
    }
    \label{fig:current_direction}
\end{figure}
As a final check of the impact of the crystal symmetries on the spin-orbit torques, we repeated our measurements and analysis for interchanged current and Hall voltage directions.
The results of these measurements are shown in different colours in \cref{fig:current_direction}a and b for \opfl{} torque \tauA{} and \ipdl{} torque \tauS{}, respectively.
As the change in resistance of the devices and device geometry is taken into account during the measurements and subsequent analysis, we expect the obtained torques to be independent of the current direction if the crystal symmetry does not play a role in the measured torques \cite{MacNeill2017}.
For the \opfl{} torque \tauA{} (\cref{fig:current_direction}a) this indeed is the case: the measured torques are virtually identical throughout the measured temperature range.
However, for the \ipdl{} torque \tauS{} (\cref{fig:current_direction}b) a small difference occurs between the two current directions: \tauS{} measured in one of the current directions is (throughout the temperature range) approximately $1.5$ times larger than \tauS{} measured in the other current direction.
This behaviour is not exclusive to the torques measured in the \NPSPy{} device; for the \Py{} device this ratio between the torques measured for different current directions is approximately $2.3$.
Since this discrepancy occurs for both the \NPSPy{} and \Py{} devices, it suggests that it is related to a measurement artefact, most likely due to different current paths for the two configurations.
Apart from this discrepancy, a very similar behaviour is found for \tauS{}, independent of the current direction.
This agrees with the conclusion drawn earlier that the torque is not related to the specific space group of the antiferromagnetic phase of \NPS{}.

\section{Measurements of additional \NPSPy{} devices}

\begin{figure}
    \includegraphics{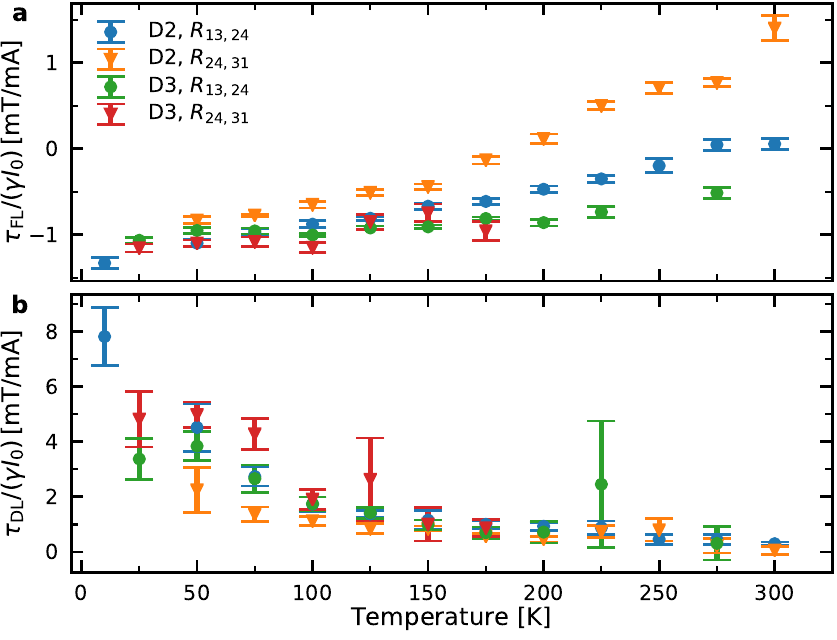}
    \caption{
        Measurements of the \opfl{} torque \tauA{} (a) and the \ipdl{} torque \tauS{} (b) for two other \NPSPy{} devices, named D2 and D3.
        The torques have been corrected for the Hall bar geometry according to \cite{Neumann2018}.
    }
    \label{fig:supportive_measurements}
\end{figure}

We performed similar measurements to the ones described in the main text for other \NPSPy{} devices, named here devices D2 and D3.
These devices were made on \NPS{} crystals with different thicknesses from the device discussed in the main text, \SI{6.34}{\nano\metre} and \SI{5.15}{\nano\metre} for D2 and D3, respectively.
\Cref{fig:supportive_measurements} shows the \opfl{} torque \tauA{} (a) and the \ipdl{} torque \tauS{} (b), for two current directions in these additional \NPSPy{} devices.

We notice that the order of magnitude of the measured torques is consistent over the different devices, though the exact value differs from device to device.
We observe that for most of our measurements, the \opfl{} torque is negative, i.e.\ opposite to the torque expected from an Oersted-field coming from an unoxidised \ce{Al} capping layer and increases in magnitude with a decrease in temperature.
Interestingly, for device D2 with the current in one particular direction ($R_{24,13}$) we observe a positive \opfl{} at room temperature with a sign change occurring at around \SI{170}{\kelvin}.
Since this behaviour is not reproduced for a different current configuration nor for other devices, we believe this is most likely due to some inhomogeneity in the current path.

For the \ipdl{} torque, D2 and D3 show, similar to D1, an increase in torque magnitude with a decrease in temperature, for all current directions.
However, for devices D2 and D3 the increase in torque magnitude is steeper than found for device D1.
For example in device D2 the torque reaches a value of \SI{8(1)}{\milli\tesla\per\milli\ampere} at \SI{10}{\kelvin}, an eightfold increase compared to the torque at room temperature.
The measured torques and device parameters are summarized in \cref{tab:torque_table} with device D1 being the one for which the data has been used for the main text.
\Cref{tab:torque_conduct_table} summarizes the torque conductivities of the measured devices.

\begin{table}
    \centering
    \caption{
        Measured torque values (both $\tauS{}/(\gamma I_0)$ and $\tauA{}/(\gamma I_0)$ in \si{\milli\tesla\per\milli\ampere}) for the different \NPSPy{} devices at room temperature (RT; \SI{300}{\kelvin}) and at low temperature (LT; \SI{50}{\kelvin}).
        Also the dimensions of the Hall bar ($W_1$ and $W_2$ in \si{\micro\metre} as defined in Fig. 1 of the main paper), the thickness of the \NPS{} flake $t_\mathrm{NPS}$ (in \si{\nano\metre}), and the sheet resistance $R_\mathrm{S}^\mathrm{RT}$ (in \si{\ohm\per\sq}) at room temperature are given.
        The torques have been corrected for the Hall bar geometry according to \cite{Neumann2018}.
    }
    \label{tab:torque_table}
    \begin{tabular}{
                |c
                |S[table-format = 1.1(0)]
                |S[table-format = 1.1(0)]
                |S[table-format = 1.1(0)]|
                |S[table-format = 3.1(0)]
                |S[table-format = 1.4(1)]
                |S[table-format = 2.4(1)]|
                |S[table-format = 2.3(1)]
                |S[table-format = 1.2(1)]|
            }  
        \toprule
        {Device} & {$W_1$} & {$W_2$} & {$t_\mathrm{NPS}$} & {$R_\mathrm{S}^\mathrm{RT}$} &
            {$\tau_\mathrm{FL}^\mathrm{RT}/(\gamma I_0)$} & {$\tau_\mathrm{FL}^\mathrm{LT}/(\gamma I_0)$} &
            {$\tau_\mathrm{DL}^\mathrm{RT}/(\gamma I_0)$} & {$\tau_\mathrm{DL}^\mathrm{LT}/(\gamma I_0)$} \\
        \midrule
              D1 &   3.0 &    2.5 &   3.1 &  140.7 &   0.079(9) &   0.140(2) &     1.0(1) &   1.4(1) \\
              D2 &   3.0 &    2.0 &   6.3 &  151.8 &    0.06(6) &   -1.10(4) &    0.30(6) &   4.5(9) \\
              D3 &   3.0 &    2.0 &   5.2 &  144.3 &       {--} &   -0.95(4) &       {--} &   3.8(5) \\
         \PtPy{} &   5.0 &    3.0 &  {--} &   15.1 &   0.231(1) &  0.2540(7) &    0.22(1) &  0.23(3) \\
           \Py{} &   5.0 &    3.0 &  {--} &  112.0 &  0.0203(2) &       {--} &  -0.018(2) &     {--} \\
        \bottomrule
    \end{tabular}
\end{table}

\begin{table}
    \centering
    \caption{
        The torque conductivities \sigmaA{} and \sigmaS{} [in \si{10^{5} (\frac{\hbar}{2e}) \per(\ohm\meter)}] is given at both room temperature (RT, \SI{300}{\kelvin} and low temperature (LT, \SI{50}{\kelvin}), assuming a saturation magnetization of $\mu_0 M_\mathrm{S} = \SI{0.7}{\tesla}$.
        Also the number of layers of the \NPS{} flake $n_\mathrm{NPS}$ \cite{Kuo2016}, and the sheet resistance $R_\mathrm{S}^\mathrm{RT}$ (in \si{\ohm\per\sq}) at room temperature are given.
    }
    \label{tab:torque_conduct_table}
    \begin{tabular}{
                |c
                |S[table-format = 1.0(0)]|
                |c
                |S[table-format = 1.4(1)]
                |S[table-format = 2.3(1)]|
                |S[table-format = 2.3(1)]
                |S[table-format = 2.1(1)]|
            }  
        \toprule
        {Device} & {$n_\mathrm{NPS}$} & {$R_\mathrm{S}^\mathrm{RT}$} &
            {$\sigma_\mathrm{FL}^\mathrm{RT}$} & {$\sigma_\mathrm{FL}^\mathrm{LT}$} &
            {$\sigma_\mathrm{DL}^\mathrm{RT}$} & {$\sigma_\mathrm{DL}^\mathrm{LT}$} \\
        \midrule
              D1 &     4 &  140.7 &    0.17(2) &   0.319(4) &     2.2(3) &     3.2(3) \\
              D2 &     9 &  151.8 &     0.1(1) &   -2.32(8) &     0.6(1) &      10(2) \\
              D3 &     7 &  144.3 &       {--} &   -2.01(8) &       {--} &       8(1) \\
         \PtPy{} &  {--} &   15.1 &    7.76(3) &    8.99(3) &     7.3(3) &       8(1) \\
           \Py{} &  {--} &  112.0 &  0.0921(9) &       {--} &  -0.079(9) &       {--} \\
        \bottomrule
    \end{tabular}
\end{table}

\section{Temperature dependence of the anomalous Hall and planar Hall resistances}

\begin{figure}[ht]
    \includegraphics{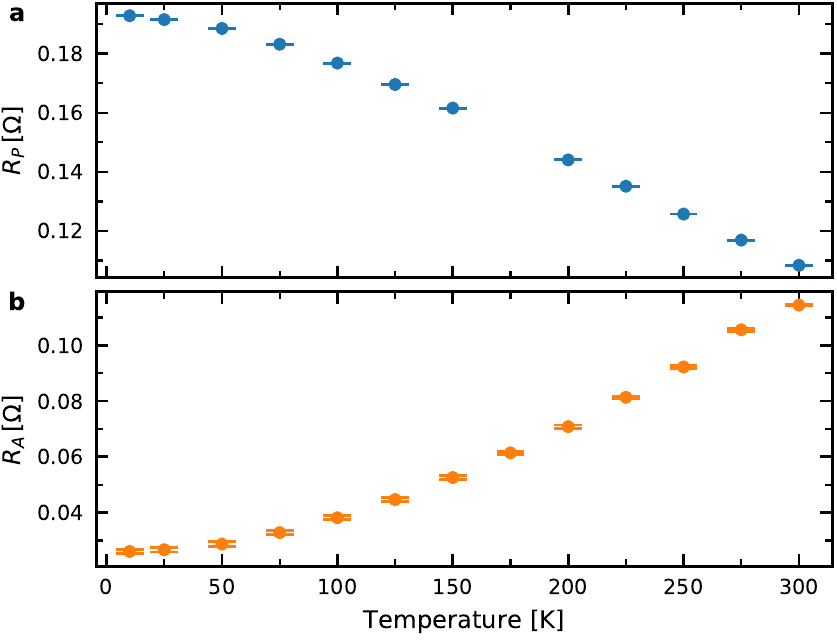}
    \caption{
        Planar Hall resistance $R_\mathrm{P}$ (a) and anomalous Hall resistance $R_\mathrm{A}$ (b) as a function of temperature.
        For each temperature, the Hall resistances are determined from the average of the measured Hall resistances at that temperature.
    }
    \label{fig:Rhe_vs_T}
\end{figure}

As a confirmation of the temperature dependence of the measured torques in the main text, the temperature dependence of both the planar Hall resistance $R_\mathrm{P}$ and the anomalous Hall resistance $R_\mathrm{A}$ were investigated.
In \cref{fig:Rhe_vs_T} these temperature dependencies are shown.
We observe a variation of both $R_\mathrm{P}$ and $R_\mathrm{A}$ as a function of temperature, which is also taken into account in our analysis for the extraction of the SOTs.

\section{Anisotropy and exchange bias in \NPSPy{} devices}
\label{sec:no_eb_an}
\begin{figure}
    \includegraphics{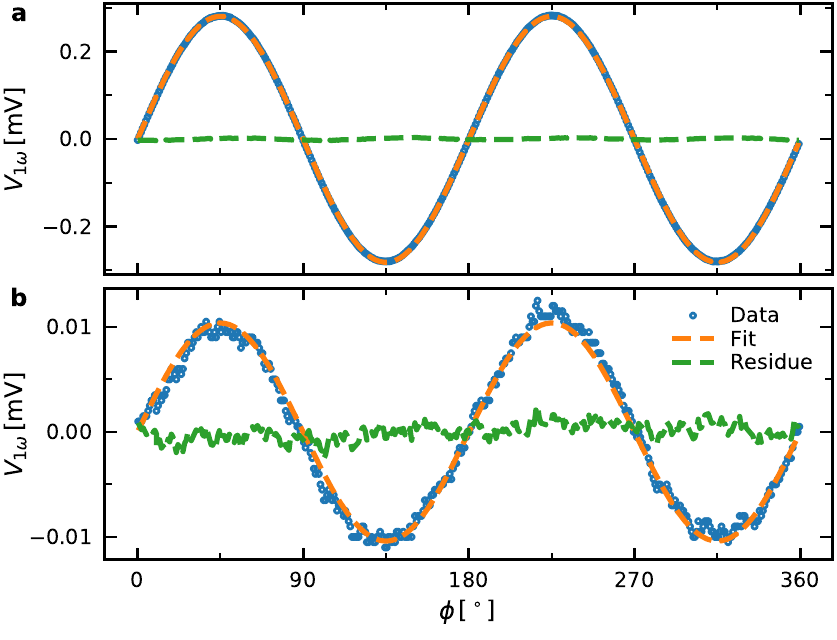}
    \caption{
        Fit and residue of fit of the first harmonic Hall voltage in a \NPSPy{} device.
        This is the same data as shown in Fig. 1 of the main text.
        For clarity purposes the residue of the fit is enlarged by a factor 10.
    }
    \label{fig:V1H_Residue}
\end{figure}

During the analysis of both the first and second harmonic Hall measurements, we assumed a lack of in-plane anisotropy or exchange bias in our devices.
Here we investigate if these assumptions are indeed valid.
\cref{fig:V1H_Residue}a shows the first harmonic Hall voltage $V_{1\omega}$ of the \NPSPy{} device D1, discussed in the main text, along with a fit of Eq. 1 of the main text and the residue of the fit.
Using this equation to fit the first harmonic Hall voltage implicitly assumes the absence of anisotropy; an additional anisotropy term is expected to cause a periodic addition to the first harmonic Hall voltage and would therefore show up in the residue of the fit.
We indeed find a small periodic component in the fit residue, indicating a very small anisotropy.
However, as this periodic component is approximately $2$ orders of magnitude smaller than the total first harmonic Hall voltage, we can safely assume that this anisotropy is sufficiently small to be disregarded in further analysis.

Similarly, we investigate if the absence of exchange bias in the \Py{} layer due to the antiferromagnetic \NPS{}.
\Cref{fig:V1H_Residue}b shows the first harmonic Hall voltage of the \NPSPy{} device after field-cooling the sample to \SI{25}{\kelvin} in a magnetic field of \SI{550}{\milli\tesla}.
In the residue of the fit using Eq.1 of the main text to the data, again a small periodic component can be found, indicating the possible presence of a small exchange bias due to the \NPS{} layer.
However, the periodic component is small compared to both the noise on the residue and the full amplitude of the first harmonic Hall voltage.
Moreover, inducing this exchange bias requires field-cooling the devices, while for the actual measurements in the main text the samples were cooled without applying a magnetic field.
Therefore, we can safely assume no exchange bias for the analysis of the measurements.

\pagebreak
\section{Fitting an out-of-plane damping-like torque}
\label{sec:fit_tauB}
\begin{figure}
    \includegraphics{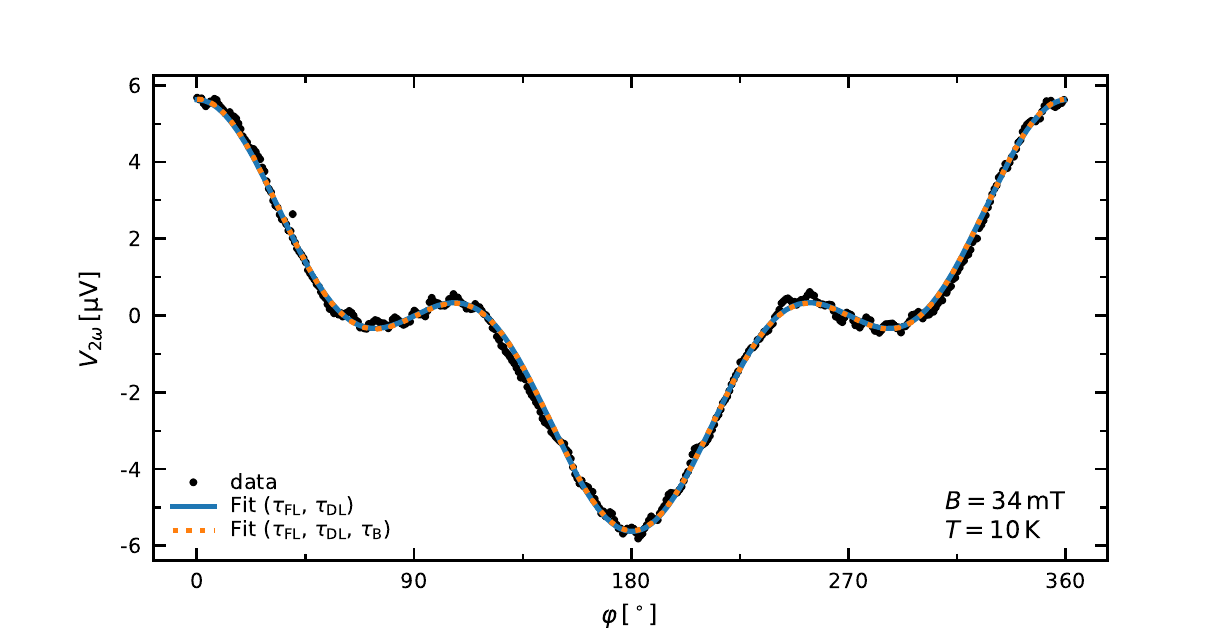}
    \caption{
        Second-harmonic Hall measurement, taken at $T = \SI{10}{\kelvin}$ and $B = \SI{34}{\milli\tesla}$, on device D1.
        The lines are two fits, excluding (blue solid line) and including (orange dashed line) contributions from the out-of-plane damping-like torque $\tau_\mathrm{B}$.
    }
    \label{fig:fit_tauB}
\end{figure}

In order to investigate the presence of another type of SOT, the out-of-plane damping-like torque ($\tau_\mathrm{B}$) we have a closer look at fitting the second harmonic Hall measurement in \cref{fig:fit_tauB}.
This measurement is taken at low temperature (\SI{10}{\kelvin}) and low field (\SI{34}{\milli\tesla}) as the contribution of the in-plane damping-like torque is expected to be strongest.
For investigating the presence of this torque we compare two fits to the second harmonic Hall voltage $V_\mathrm{H}^{2\omega}$ with a modified version of Eq. 2 of the main text \cite{MacNeill2017b}:
\begin{equation}
    \label{eq:VH2_tauB}
    V_\mathrm{H}^{2\omega} = 
        -  I_0 (R_\mathrm{FL} \sin\varphi + R_\mathrm{B}) \cos2\varphi
        -  \frac{1}{2} I_0 R_\mathrm{DL} \sin\varphi,
\end{equation}
where $I_0$ is the applied current, $\varphi$ the angle of current with the magnetic field direction, and $R_\mathrm{FL}$, $R_\mathrm{DL}$, and $R_\mathrm{B}$ is the coefficient (with units \si{\ohm}) proportional to \tauA{}, \tauS{} and the out-of-plane damping-like torque $\tau_\mathrm{B}$, respectively.
By comparing a fit where $R_\mathrm{B}$ is fixed to zero with a fit where $R_\mathrm{B}$ is varied, we can determine if the $R_\mathrm{B}$ has a significant contribution and hence if $\tau_\mathrm{B}$ is present.

Both of these fits are shown in \cref{fig:fit_tauB}.
It is clear that adding $R_\mathrm{B}$ does not have a significant impact on the line-shape of the fit; the fits overlap both with each other and with the data.
The values resulting of the fits also suggest that $R_\mathrm{B}$ does not have a significant influence.
When varied, $R_\mathrm{B}$ is found to be \SI{6(4)e-06}{\ohm}, compared to \SI{1.42(1)e-03}{\ohm} and \SI{1.369(8)e-03}{\ohm} for $R_\mathrm{FL}$ and $R_\mathrm{DL}$, respectively.
Moreover, the values of both $R_\mathrm{FL}$ and $R_\mathrm{DL}$ do not vary (up to \SI{1e-3}{\percent}, well within the margin of error) when $R_\mathrm{B}$ is fixed to zero instead of varied.
Hence, we conclude that $\tau_\mathrm{B}$ is not present in these devices.

\bibliography{Bibliography}